# Report on Neural-like Criticality in Ag-based Nanoparticle Networks

*Blessing Adejube[1], Jamie Steel[2], Joshua Mallinson[2], Mariia Protsak[3], Daniil Nikitin[3], Thomas Strunskus[1,4], Andrei Choukourov[3], Franz Faupel[1,4], Simon Anthony Brown[2], and Alexander Vahl[1,4,5]\**

[1] Department of Materials Science – Chair for Multicomponent Materials, Faculty of Engineering, Kiel University, Kaiserstraße 2, D-24143 Kiel, Germany

[2] The MacDiarmid Institute for Advanced Materials and Nanotechnology, School of Physical and Chemical Sciences, University of Canterbury, Christchurch, New Zealand

[3] Charles University, Faculty of Mathematics and Physics, Department of Macromolecular Physics, V Holešovičkách 2, 180 00 Prague, Czech Republic

[4] Kiel Nano Surface and Interface Science KiNSIS, Kiel University, Christian-Albrechts-Platz 4, D-24118 Kiel, Germany

[5] Leibniz Institute for Plasma Science and Technology, Felix-Hausdorff-Straße 2, 17489 Greifswald, Germany

correspondence to: alva@tf.uni-kiel.de

Abstract

Emulating the neural-like information processing dynamics of the brain provides a time and energy efficient approach for solving complex problems. While the majority of neuromorphic hardware currently developed rely on large arrays of highly organized building units, such as in rigid crossbar architectures, in biological neuron assemblies make use of dynamic transitions within highly parallel, reconfigurable connection schemes. Neuroscience suggests that efficiency of information processing in the brain rely on dynamic interactions and signal propagations which are self-tuned and non-rigid. Brain-like dynamic and avalanche criticality have already been found in a variety of self-organized networks of nanoobjects, such as nanoparticles (NP) or nanowires. Here we report on the dynamics of the electrical spiking signals from Ag-based self-organized nanoparticle networks (NPNs) at the example of monometallic Ag NPNs, bimetallic AgAu alloy NPNs and composite Ag/ZrN NPNs, which combine two distinct NP species. We present time series recordings of the resistive switching responses in each network and showcase the determination of switching events as well as the evaluation of avalanche criticality. In each case, for Ag NPN, AgAu NPN and Ag/ZrN NPN, the agreement of three independently derived estimates of the characteristic exponent provides evidence for avalanche criticality. The study shows that Ag-based NPNs offer a broad range of versatility for integration purposes into physical computing systems without destroying their critical dynamics, as the composition of these NPNs can be modified to suit specific requirements for integration.

## 1. Introduction

Material systems that show properties similar to bio-neural networks are undergoing active development[1–4]. The findings from neuroscience that the brain operates within a critical regime [5–8,8,9], has motivated a lot of research interests to explore criticality and critical properties in artificial networks [10–14]. This helps with adaptation and cognition in the brain [6,15], and is considered beneficial for imparting adaptation into artificial systems, solving complex tasks using unconventional computing approaches such as reservoir computing (RC) [10,16–18]. Networks of percolating NPs were shown to exhibit brain-like avalanche dynamics and neural-like spiking dynamics like self-organization, scale-invariance, long-range-temporal correlation (LRTC) and stochasticity [10,11,13,19]. These nanogranular networks have also been used for RC to perform certain benchmark tasks such as temporal classification, chaotic time series prediction, speech processing, and nonlinear transformation [16–18]. In NPNs at percolation, in between insulating and conducting states, multiple metastable junctions are present. Within these metastable nanojunctions, a switching event (change in resistance or conductance) is capable of giving rise to further events across various nanojunctions throughout the network. In this manner an avalanche can be generated and the propagation of an avalanche is referred to as a percolation phenomenon [8,10,20]. Critical dynamics in self-organized networks have been established and a previous study showed that for silver (Ag) NPNs, covering with a matrix does not destroy the power-law scaling behavior of its avalanche [1,11,19]. However, a more detailed understanding of the robustness of the avalanche criticality (especially for the commonly studied Ag), to nanoscale modification is still lacking.

Here, we report on collective resistive switching properties in three Ag-based NPNs: a silver-gold alloy (AgAu) NPN, a composite NPN from silver NPs and zirconium nitride NPs (Ag/ZrN), and a monometallic Ag NPN. The present focus is not comparison between the different NPNs but to assess the prevalence of avalanche criticality in each of the Ag-based

NPNs. Using the approach as in [10], the presence of avalanche criticality was demonstrated for each NPN. By this study, we show that the Ag-based NPNs can be modified to suit specific requirements for integration with other computing components, without breaking its collective resistive switching properties and critical dynamics.

## 2. Results and discussion

AgAu-, Ag/ZrN-, and Ag NPNs were deposited until the onset of conduction to create a network at the transition regime between conduction and insulation. All NPNs under a 3 V bias show current spiking response within the mA regime showing that the NPNs are not fully conducting. To create NPNs that can show critical dynamics, deposition was halted at the onset of conduction (mA regime for Ag-, Ag/ZrN and µA regime for AgAu). As a result, the NPN is at a critical state where a change in the conductance (or resistance) at a nanojunction can trigger subsequent events leading to a cascade or bursts of events as described in [10,11,13].

### 2.1 Demonstration of avalanche criticality and event correlation

A characteristic trait of critical systems is that their avalanche sizes (S) and durations (T) have a probability density function (PDF) that can be described in form of a power law, $P(S) \sim S^{-\tau}$ and $P(S) \sim S^{-\alpha}$, as described in [10,21]. In these systems, the avalanche mean sizes are related to their durations by a characteristic exponent $1/\sigma vz$, and this is one condition for criticality. The characteristic exponent generated should be then consistent with the exponent obtained from the crackling noise relationship which is another condition for criticality [21]

$$\frac{1}{\sigma vz} = \frac{\alpha - 1}{\tau - 1} \qquad (1)$$

Additionally, the avalanche shapes (spatial or temporal distribution of the activity during an avalanche) should show a "collapse" onto a single curve to independently yield the characteristic scaling exponent $1/\sigma vz$. Collapse here refers to the concept of scaling such that when the avalanche profile is rescaled, for example by normalization, the avalanches should follow the same pattern and will appear to have the same shape when plotted on the rescaled axis. By achieving the data collapse, an independent estimate of the exponent can be extracted which then confirms that the events within the system are not some random telegraph noise [21–23]. These three independent estimates of the characteristic exponent should be consistent for a system to be critical [21,24,25].

## 2.2 Section A: Collective resistive switching in AgAu NPN

### 2.2.1 Detection of switching events in AgAu NPN

In AgAu NPN, a time series measurement over 6000 s at a measuring frequency of 1kHz under a constant bias of 3 V between two electrodes, showed a complex current spiking response (Figure 1a), with periods of silence and bursts in activity. To detect the relevant switching events, a threshold of $\Delta G = 0.02$ $G_0$ (where $G_0 = 2e^2/h$), was used, and all $\Delta G$ below the threshold (red line in Figure 1b) were regarded as noise.

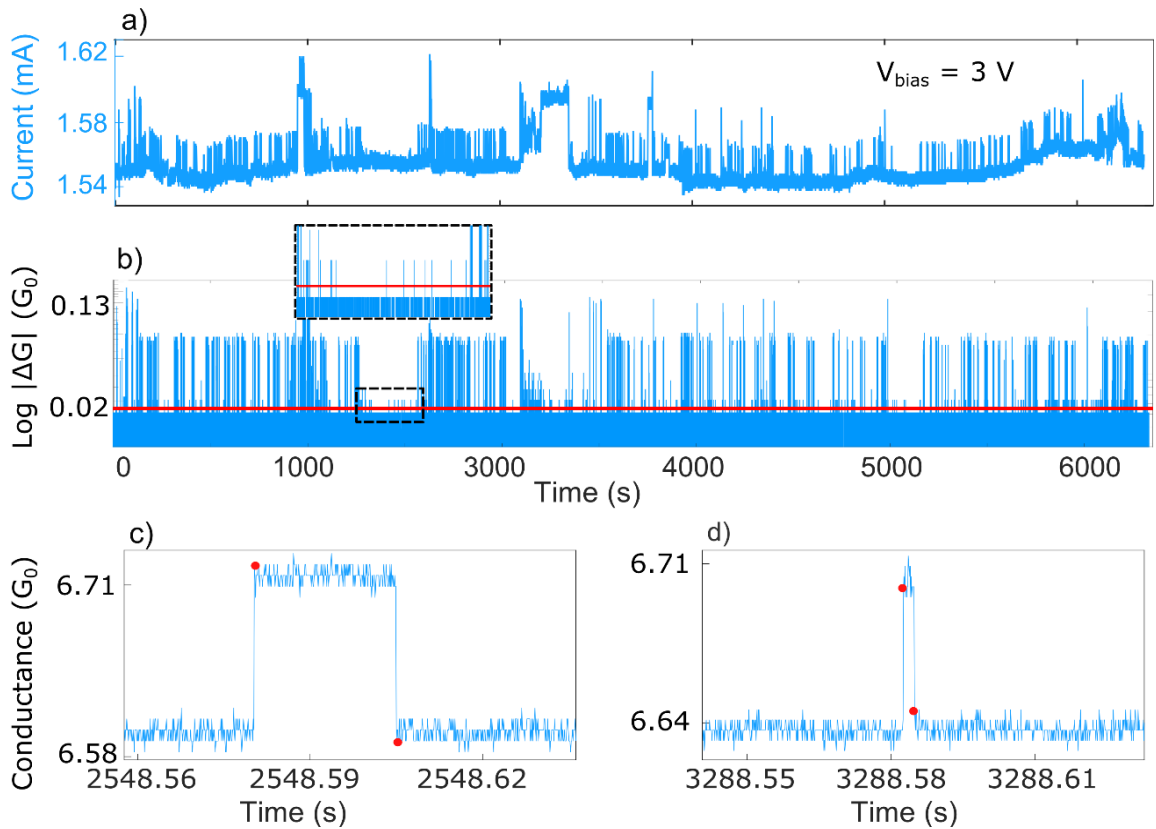

Figure 1: Current time-series measurement, spiking signals and thresholding of spiking signals in AgAu NPN. In a), the time-series recording at a sampling rate of under a constant bias of 3 V and the log of absolute conductance changes shown in b) with the red line indicating the threshold. The zoom in on the dotted square as shown in the inset, gives a clearer view of the threshold line. The threshold was sufficient to capture the signals as the switching events were quite sharp as seen in c) and d). Red markers indicate the captured events.

As shown by the red line in Figure 1b and the zoom in of the marked area, this threshold was sufficient to eliminate noise and capture the actual spiking signals as shown in Figure 1c and d. In each case, two switching events are detected, which are marked by a red dot in Figure 1c and d. The thresholded events were then used for statistical evaluation and demonstration of avalanche criticality.

**2.2.2 Avalanche criticality and event correlation in AgAu NPN**

Figure 2a – i presents the result of the characterization of avalanche dynamics in the AgAu NPN. As shown in Figure 2a, the distribution of ΔG follows a power law and so does the distribution of inter-event intervals (IEI) (Figure 2b). The probability density function of the IEI follows a power law dependency for over six orders of magnitude, which indicates collective resistive switching with correlation between events. The autocorrelation function (ACF) is described by a power law (Figure 2c), also showing events are correlated. In Figure 2d and e, the avalanche size and durations are depicted, which also indicate power law dependencies. The relation between the mean avalanche size ($\langle S \rangle$) and the duration (T), is shown in Figure 2f. The mean avalanche shapes (t), exhibit a collapse as presented in Figure 2g and h. The exponent of the crackling noise relationship was 1.4 with a margin error of ± 0.2. Nevertheless, in Figure 2i, the characteristic exponents of the independent estimates obtained from the shape collapse (cyan), crackling noise relationship (green) and that of the mean avalanche sizes (blue) were consistent and overlap within the error margin. The power law exponent of the avalanche size $\tau$, (red) is also shown in Figure 2i. The demonstration that all three independent estimates of $1/\sigma vz$ are consistent, presents very strong evidence for criticality in the AgAu NPN.

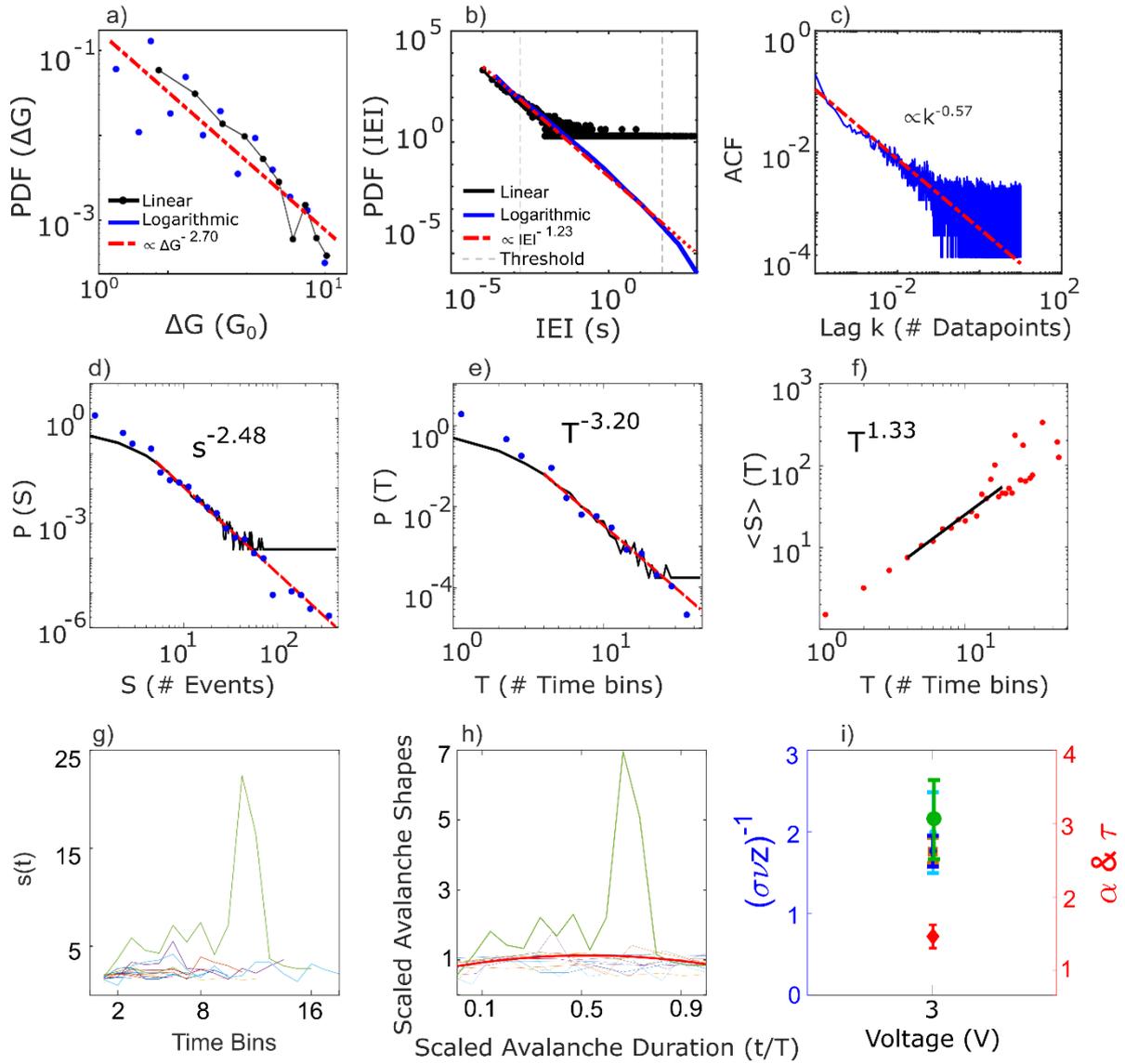

Figure 2: Demonstration of avalanche criticality in AgAu NPN. PDF of the ΔG in a) follow a power law behavior, and the distribution of the IEI in b), follows a power law for over six orders of magnitude which indicates that the events are correlated. Temporal correlation between the events (blue) is shown by the ACF in c), which follows the power law over several orders of magnitude and together with the IEI, provides strong evidence for temporal self-similarity. In d) and e), linear bin sizes are presented in black, and heavy tailed distribution can be observed from the logarithmic bin sizes in blue. The PDF of the avalanche sizes d) and avalanche duration, e) both follow a power law, with critical exponents of $\tau = 2.48 \pm 0.01$ and $\alpha = 3.2 \pm 0.01$ respectively. In f), the ⟨S⟩ are a power law function of their duration with the characteristic exponent $1/\sigma vz = 1.33 \pm 0.08$. The mean avalanche shapes s(t) in g), are shown to exhibit shape collapse on to a single curve h). The shape collapse gives an independent estimate of the critical

exponent $1/\sigma vz = 1.4 \pm 0.2$ which then lies within the measurement error margin and satisfy the crackling relationship. In i), the estimate of the three independent determinations of $1/\sigma vz$ as obtained from the crackling relationship (green), the plotted dependence of the mean avalanche size ⟨S⟩ (T) for given durations (blue), and the avalanche shape collapse (cyan) are consistent within the measurement uncertainty. The power law exponent of the avalanche size $\tau$, (red) and duration $\alpha$, (brown) are shown.

**2.3 Section B: Collective resistive switching in Ag/ZrN NPN**

**2.3.1 Detection of switching events in Ag/ZrN NPN**

In Figure 3a, the current response of percolating Ag/ZrN NPN under 3 V bias for over 6000 s and a sampling frequency of 1 kHz is presented. A threshold of $\Delta G = 0.02\ G_0$ was used and all $\Delta G < 0.02\ G_0$ (red line in Figure 3b) were regarded as noise as shown more clearly by the zoom in. The $\Delta G$ signals in AgZrN were sharp and quite distinguishable from the noise hence the thresholding was sufficient to capture relevant data as indicated by the red markers in Figure 3c. Occasionally, the spiking events show a complex shape with gradual changes in the current response, as seen in Figure 3d. To fully capture the conduction changes in the network, the gradual changes in the network response are also treated as switching event and are considered for analysis of the collective switching behaviour.

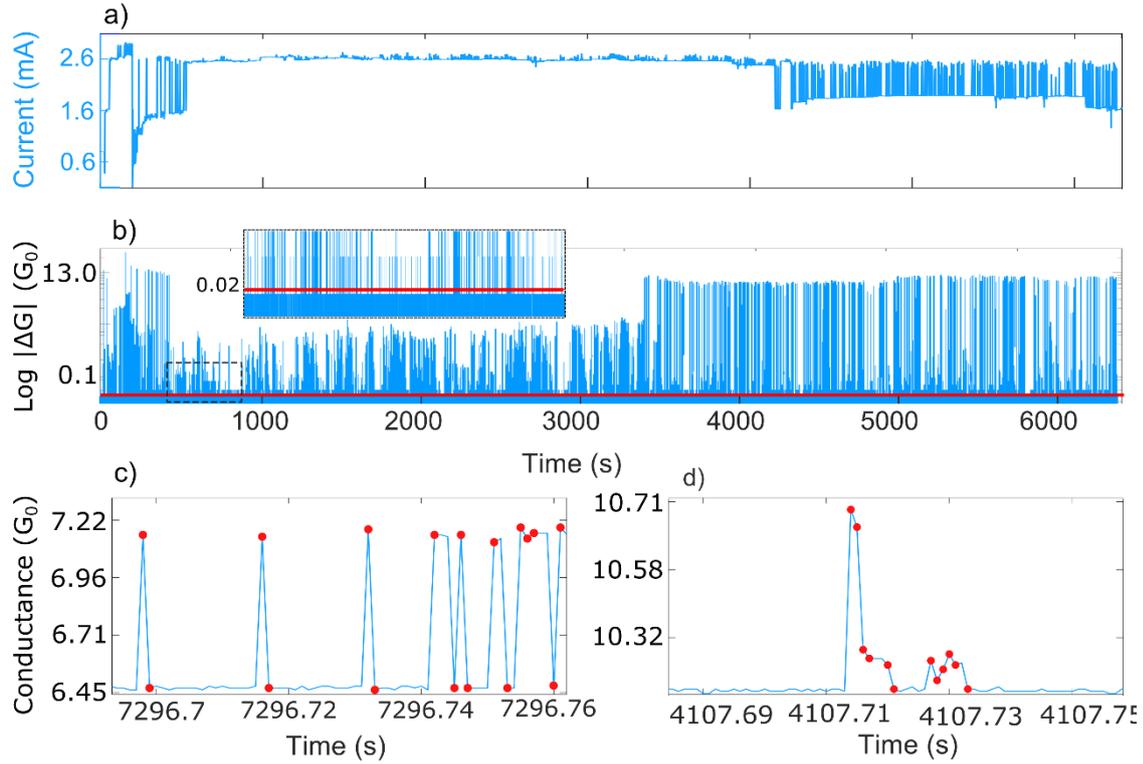

Figure 3: Measured current spikes and thresholding of spiking signals in Ag/ZrN NPN. Current response under a constant 3 V bias over time is shown in a). Threshold line for the detected absolute changes by eliminating events, ΔG < 0.02 $G_0$, is as shown in b) and in the zoom in on the dotted square for better visualization. This threshold was sufficient to capture actual spiking events (red markers) as shown in c) and d) without taking in noise.

### 2.3.2 Demonstration of avalanche criticality and event correlation in Ag/ZrN NPN

In Figure 4a – i, the analysis of criticality in the Ag/ZrN NPN shows that the PDFs of ΔG and of the IEI histograms (Figure 4a and b), are heavy-tailed and follow a power law behavior over several orders of magnitude. This indicated that the switching events are correlated which is then consistent with the power law behavior of the ACF in Figure 4c. In Figure 4d and e, the avalanche sizes and durations are also shown to follow well defined power law distributions. The mean avalanche size is shown to depend on the duration and the crackling relationship is satisfied (Figure 4f). The avalanches exhibit shape collapse (Figure 4g) onto a single curve (Figure 4h). The three independent estimates of the characteristic exponent $1/\sigma vz$, all overlap and are consistent (Figure 4i) providing evidence for criticality in the Ag/ZrN NPN.

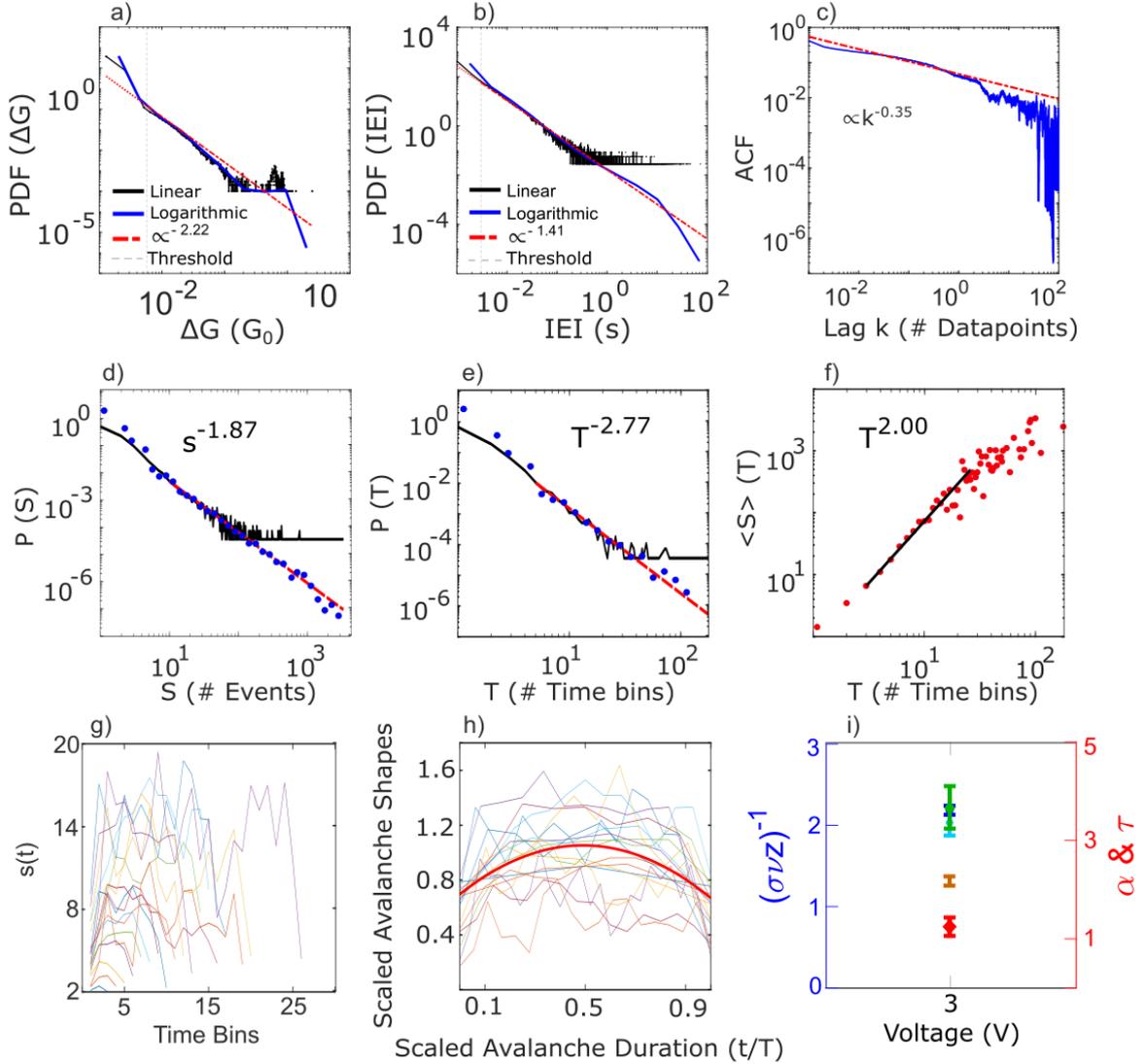

Figure 4: Demonstration of avalanche criticality in Ag/ZrN NPN. In a), a very clear power law behavior of ΔG for about four orders of magnitude and it can be seen that ΔG exhibits a heavy-tailed distribution (blue) consistent with spatial self-similarity. In b), the histogram of the IEI is also heavy-tailed and follow the power law distribution for over five orders of magnitude showing a scale free dynamics. Temporal correlation of the switching activity (blue) is shown by the power law of the ACF in c). Together with the IEI, it provides strong evidence for temporal self-similarity. In d) and e), linear bin sizes are presented in black, and heavy tailed distribution can be observed from the logarithmic bin sizes in blue. The PDF of the avalanche sizes d) and avalanche duration, e) both follow a power law (providing evidence of temporal correlation), with critical exponents of $\tau = 1.87 \pm 0.1$ and $\alpha = 2.77 \pm 0.1$ respectively. In f), the average size of an avalanche, ⟨S⟩, scales with its duration (T) defined by a power-law relationship characterized by the critical exponent $1/\sigma vz = 2.0 \pm 0.04$. The mean avalanche shapes s(t), in g) are shown to exhibit shape collapse on to a single curve h). The shape collapse yields an independent estimate of the critical exponent $1/\sigma vz = 1.9 \pm 0.1$. This then satisfies the crackling relationship as both values (mean avalanche size given duration and crackling relationship) are consistent. In i), the estimate of the three independent determinations of $1/\sigma vz$ as obtained from the crackling relationship (green), the plotted dependence of the mean avalanche size ⟨S⟩ (T) for given durations (blue), and the avalanche shape collapse (cyan) are consistent within the measurement uncertainty. The power law exponent of the avalanche size $\tau$, (red) and duration $\alpha$, (brown) are also presented. The consistency of all three independent estimates of $1/\sigma vz$ provides a strong evidence for criticality in Ag/ZrN NPN.

**2.4 Part C: Collective resistive switching in Ag NPN**

**2.4.1 Detection of switching events in Ag NPN**

In Figure 5a, the time series recording of the current response under an constant applied bias of 3 V shows periods of quiescence and burst in activity. Different from an earlier study where measurement was taken over a longer time [11], here, a high measurement frequency of 100 kHz was used to measure over a period of 60 s which provided sufficient data required for the statistical evaluation. To determine the actual switching events, which are characterized by changes in the conductance, a thresholding step as shown in Figure 5b – d, was applied. Events corresponding to a conductance change of less than 0.01 $G_0$ were neglected. The zoom in (Figure 5b) showed that the threshold was at the border of actual events and noise. It can be observed from Figure 5c and d that there are multiple consecutive and gradual switching events. The non-rectangular shape of the spiking events with gradual changes in the conductance poses a dilemma for the identification and assignment of individual spiking events, which has further consequences for the analysis of avalanche criticality. Two approaches can be followed:

A) Considering every conductance change above the threshold as an individual switching event: Conventional thresholding with a fixed absolute threshold can be applied. All conductance change events with a change greater than the conductance thresholds are accounted for and considered as individual switching events. In consequence, for a gradual rise or fall in conductance within consecutive conductance change events, a large number of switching events with very short inter-event-intervals will be recorded and the overall distribution of the absolute conductance changes will be shifted towards lower values.

B): Summarizing gradual conductance change events into distinct spiking events: Assuming that consecutive, gradual conductance change events trace back to a single spiking event in the network, the application of a constant threshold would not be applicable as it would pose the risk of underrepresenting spiking events with small slopes, which in turn would result in a shift of the IEI distribution towards larger values.

For this study, we followed approach (A) and applied a constant threshold of 0.01 $G_0$ throughout the dataset. Consequently, a large number of individual switching events was identified (red dots in Figure 5c and d. However, some complex changes which were made up of gradual changes that are very small and close to the noise level may occasionally be missed, as shown by the black circle in Figure 5d. For future analysis of such complex conduction change profiles, there is a demand for more adaptable and configurable thresholding processes.

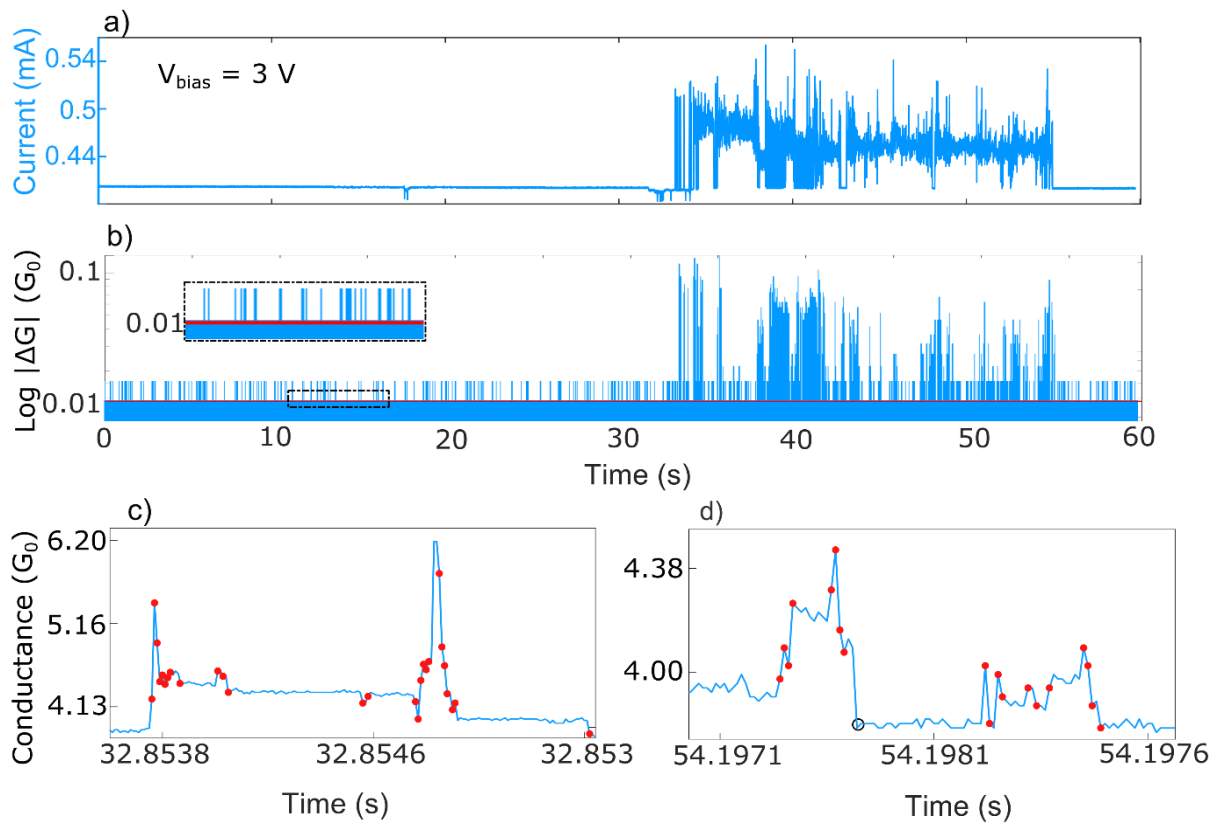
12

Figure 5: Measured spiking signals and thresholding in Ag NPN. In a), the raw measurement signal at a sampling rate of a hundred thousand points per second, under a constant bias of 3 V is shown over a period of 60 s. To detect the noise limit, the absolute difference in conductance is thresholded as shown by the red line in b) and in the zoom in. Conductance changes or events above the red line in b) ($\Delta G > 0.01$ $G_0$), are considered as events. In view of the thresholding, c) and d) shows how the events are selected. Each red marker in c) and d) represent a detected event, and as shown, although few events might be missed (black circle in d), the detected events are such that they are above the general noise level.

**2.4.2 Demonstration of avalanche criticality and event correlation in Ag NPN**

In 6a, the distribution of ΔG in the Ag NPN shows a heavy tail (i.e. the probability of extreme events, or events with large values, decay with a power law) which is also an indication of self-similarity in percolating systems. Histograms of the inter-event-interval (IEI) in Figure 6b and the auto correlation function in Figure 6c follow a power law behavior. To demonstrate criticality, the avalanche sizes, Figure 6d, and avalanche durations, Figure 6e, are described by a power law for both linear and log bins. Figure 6f shows that the mean sizes of the avalanches are dependent on the durations as required for critical systems. The values of the critical exponents $\tau, \alpha$, and $1/\sigma\nu z$, satisfied the crackling relationship (Figure 6f). Also, the exponents from the crackling noise relationship in Figure 6f and that of the shape collapse in Figure 6g and h were within the margin error (Figure 6i) and they independently yielded characteristics exponents which were consistent which satisfy the requirement for criticality.

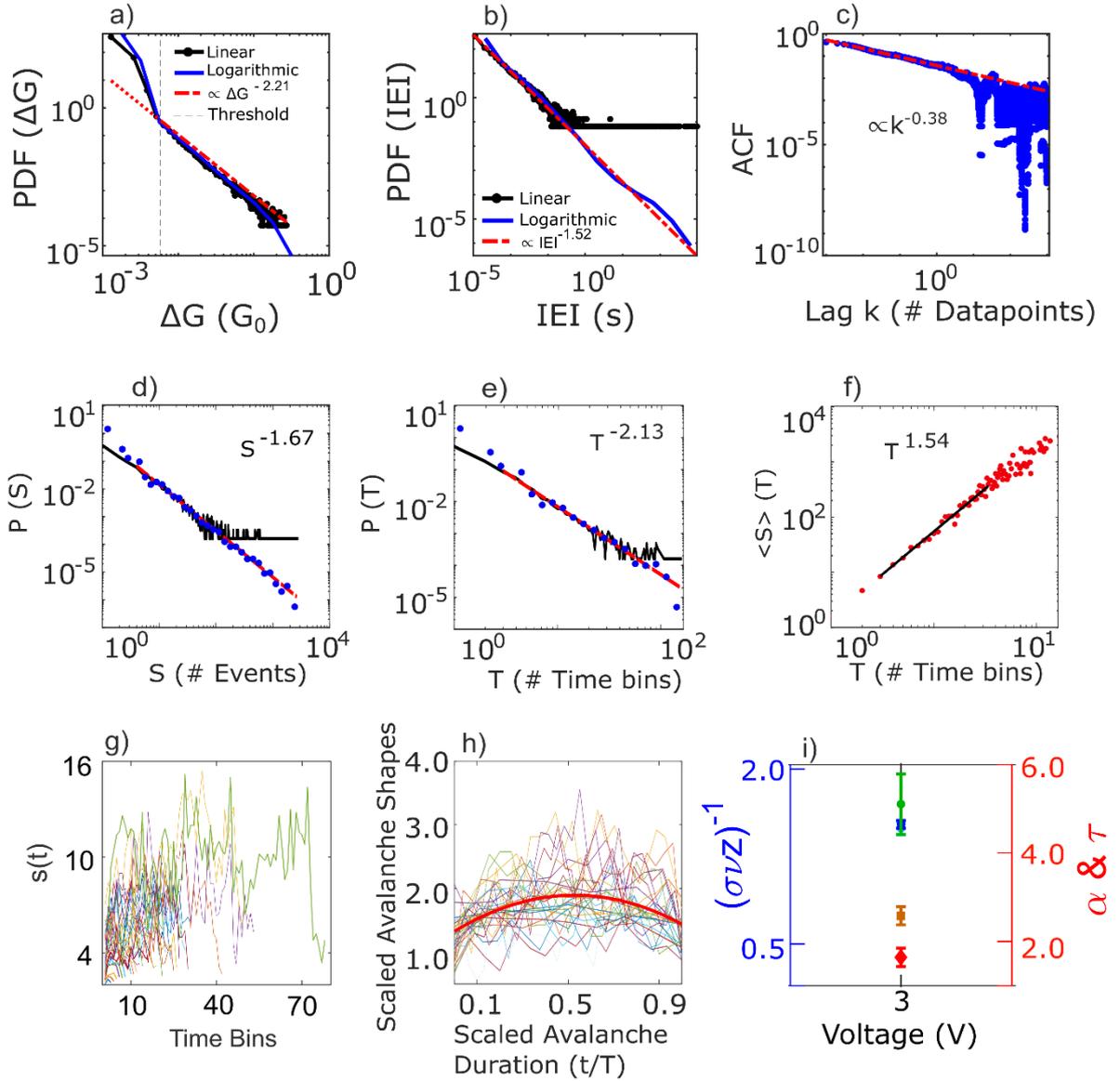

Figure 6: Demonstration of avalanche criticality in Ag NPN. In a), the probability density function, PDF, of the conductance changes, ΔG, (arising from the switching activities within the NPN) follows a power law behavior for over four orders of magnitude. The exponent of ΔG is given as 2.21 with a heavy tailed power law behavior indicating spatial self-similarity. The inter-event-interval (IEI) in b), with an exponent of 1.52, shows a power law behavior which indicates that the events are correlated. Temporal correlation of the switching activity (blue) is shown by the auto correlation function (ACF) in c), which follows the power law over three orders of magnitude. Together with the IEI, it provides strong evidence for temporal self-similarity. In d) and e), linear bin sizes are presented in black, and heavy tailed distribution can be observed from the logarithmic bin sizes in blue. The PDF of the avalanche sizes d) and avalanche duration, e) both follow a power law, with critical exponents of $\tau = 1.67 \pm 0.01$ and $\alpha = 2.13 \pm 0.01$ respectively. In f), the dependence of the mean avalanche sizes ⟨S⟩ on avalanche durations (T), is shown. It is shown in f) that the ⟨S⟩ are a power law function of their duration with the characteristic exponent $1/\sigma\nu z = 1.5 \pm 0.02$. The mean avalanche shapes s(t) in g) are shown to exhibit shape collapse on to a single curve h). The shape collapse gives

an independent estimate of the critical exponent $1/\sigma\nu z = 1.5 \pm 0.05$ which then satisfies the crackling relationship. In i), the estimate of the three independent determinations of $1/\sigma\nu z$ as obtained from the crackling relationship (green), the plotted dependence between the mean avalanche size ⟨S⟩ (T) for given durations (blue), and the avalanche shape collapse (cyan) are consistent within the measurement uncertainty. The power law exponent of the avalanche size $\tau$, (red) and duration $\alpha$, (brown) are presented and as is immediately clear, the critical exponents exhibit a very narrow margin of error.

## 3. Conclusion

This study reports on the collective resistive switching under constant applied DC bias for Ag-based NPNs and showcases the characterization of avalanche criticality for Ag, AgAu and Ag/ZrN NPNs. Similar to the avalanche critical dynamics observed in the neurons, it is demonstrated that the electrical spiking signals in time series recordings of percolating Ag-based NPNs show features of collective resistive switching and avalanche criticality and outline that the resistive switching events differ from random telegraph noise. The observation of power law scaling of switching amplitude, inter-event-intervals and the satisfaction of crackling noise relation is in agreement with earlier reports on Ag NPNs (with and without covering by $SiO_xN_y$)[11]. Three independently derived estimates of the characteristic exponent $1/\sigma\nu z$ are in reasonable agreement, which implies the presence of neural-like avalanche criticality in each of the showcased Ag-based NPNs. Consequently, Ag-based NPNs are a promising material platform for brain-inspired information processing, offering a large exploration space for modifications in composition, morphology and electrical properties, while being capable of showing collective resistive switching with avalanche criticality.

## 4. Materials and methods

### 4.1 Fabrication of percolating NPNs

The Haberland gas aggregation cluster source (GAS) was used for the deposition of Ag NPs onto a $SiO_2$/Si (10 mm by 10 mm) substrate having ten Pt electrodes (80 nm), patterned via UV-lithography and lift-off, in a circular orientation with a spacing of 28 µm between neighboring electrodes. Before deposition of NPs, the chips were ultrasonically cleaned with isopropanol (10 mins) and deionized water (5 mins). A custom-made sample holder and printed circuit board (PCB) that enabled contacting of the electrodes during deposition was used to hold the chip during deposition. This holder was connected to a Keithley 2450 source measure Unit for *in situ* electrical monitoring. By this, under a bias of 3 V (which so far, does not influence the network's morphology), the current response was measured, and the deposition automatically stopped at the onset of conduction. Detection of conduction was done by the system, depending on the resistance value given by the user, and a python code that controls the shutter. The deposition times were 4.5 min, 2.0 min, and ~9 min for Ag, AgAu and Ag/ZrN NPNs, respectively. Figure SI 1 shows the change in current (transition to percolation) of the *in situ* measurement during fabrication for all NPNs. For Ag/ZrN NPN, a pre-deposition of ZrN NPs was done as in [27,28] to have less than a monolayer (84%) of ZrN NPs base coverage. The alloy (AgAu) NPN was fabricated using a custom-made Ag target embedded with gold wires, which was used as described in [29].

## 4.2 Electrical characterization

The electrical measurements were carried out using the National Instruments multichannel data acquisition system (NI PXIe-1082 and PXIe-6378 ADC/DAC) and a custom-built sample mounting system as described in [8]. Samples were about 4 – 9 weeks old during the period of electrical characterization. During these measurements, a constant input signal was applied to one of the electrodes while the output was read out from 5 other electrodes.

## 4.3 Data analysis and power law analysis

Data analysis methods used are, in principle, like those used in neuroscience for analysis of brain tissue electrical signal [9,23,25,30] and as was done on NPNs, in [10]. The width of the time bins equals the mean IEI. Power law exponents and the fittings were done by using the maximum likelihood estimate (MLE) as shown in [10,31], and the Complexity and Criticality MATLAB toolbox by [10,31]. The power law and avalanche statistical evaluation is extensively discussed in [10,11].

## 5. Acknowledgements

Funding support by the Deutsche Forschungsgemeinschaft (DFG, German Research Foundation) – Project-ID 434434223 – SFB 1461.

B. Adejube and A. Vahl also acknowledge support from the German Academic Exchange Service (DAAD) - Project-ID: 57654440. D. Nikitin, M. Protsak, A. Choukourov, acknowledge the support via grant GACR 23–06925S from the Czech Science Foundation. D. Nikitin, M. Protsak, acknowledge the Czech-German Mobility grant: 8J23DE016 from the Ministry of Education, Youth and Sports.